\documentclass[aps,prb,showpacs,floatfix,twocolumn]{revtex4-1}

\usepackage{graphicx}
\usepackage{setspace}
\usepackage{rotating}
\usepackage{amsmath}
\usepackage{amssymb}
\usepackage{multirow}

\def\d{{\rm d}}

\begin{document}

\title{Perspective: Advances and challenges in treating van der Waals dispersion forces in density functional theory}
\author{Ji\v{r}\'\i\ Klime\v{s}}
\author{Angelos Michaelides}
\email{angelos.michaelides@ucl.ac.uk}
\affiliation{Thomas Young Centre, London Centre for Nanotechnology and Department of Chemistry, 
University College London, London, WC1E 6BT, UK}


\date{\today}

\begin{abstract}
Electron dispersion forces play a crucial role in determining the structure and properties
of biomolecules, molecular crystals and many other systems. 
However, an accurate description of dispersion 
is highly challenging, with the most widely used electronic structure technique, 
density functional theory (DFT), failing to describe them with standard approximations.
Therefore, applications of DFT to systems where dispersion is important have traditionally been 
of questionable accuracy. 
However, the last decade has seen a surge of enthusiasm in the DFT community
to tackle this problem and in so-doing to extend the applicability of DFT-based methods. 
Here we discuss, classify, 
and evaluate some of the promising schemes to emerge in recent years. A brief perspective on the outstanding
issues that remain to be resolved and some directions for future research are also provided.

\vskip 1cm
Copyright (2012) American Institute of Physics. This article may be
downloaded for personal use only. Any other use requires prior permission of the author 
and the American Institute of Physics. 
The following article appeared in J. Chem. Phys. {\bf 137}, 120901 (2012) and may be found at
\href{http://jcp.aip.org/resource/1/jcpsa6/v137/i12/p120901_s1}{http://jcp.aip.org/resource/1/jcpsa6/v137/i12/p120901\_s1}.
\end{abstract}

\maketitle

\section{Introduction}

The theoretical description of matter as well as of many chemical, physical, and biological processes 
requires accurate methods for the description of atomic and molecular-scale interactions.
Whilst there are many quantum mechanical approaches, in the past few decades Kohn-Sham density functional 
theory (DFT)\cite{hohenberg1964,kohn1965} has established itself as the theoretical method of 
choice for this task, undergoing a meteoric rise in large parts of physics, chemistry, and materials science.
The rise of DFT and its uptake in academia and industry has been widely discussed, and was perhaps illustrated most clearly 
in Burke's recent $Spotlight$ article on DFT.\cite{burke2012jcp}

\begin{figure}[h]
\centerline{
\includegraphics[width=6cm]{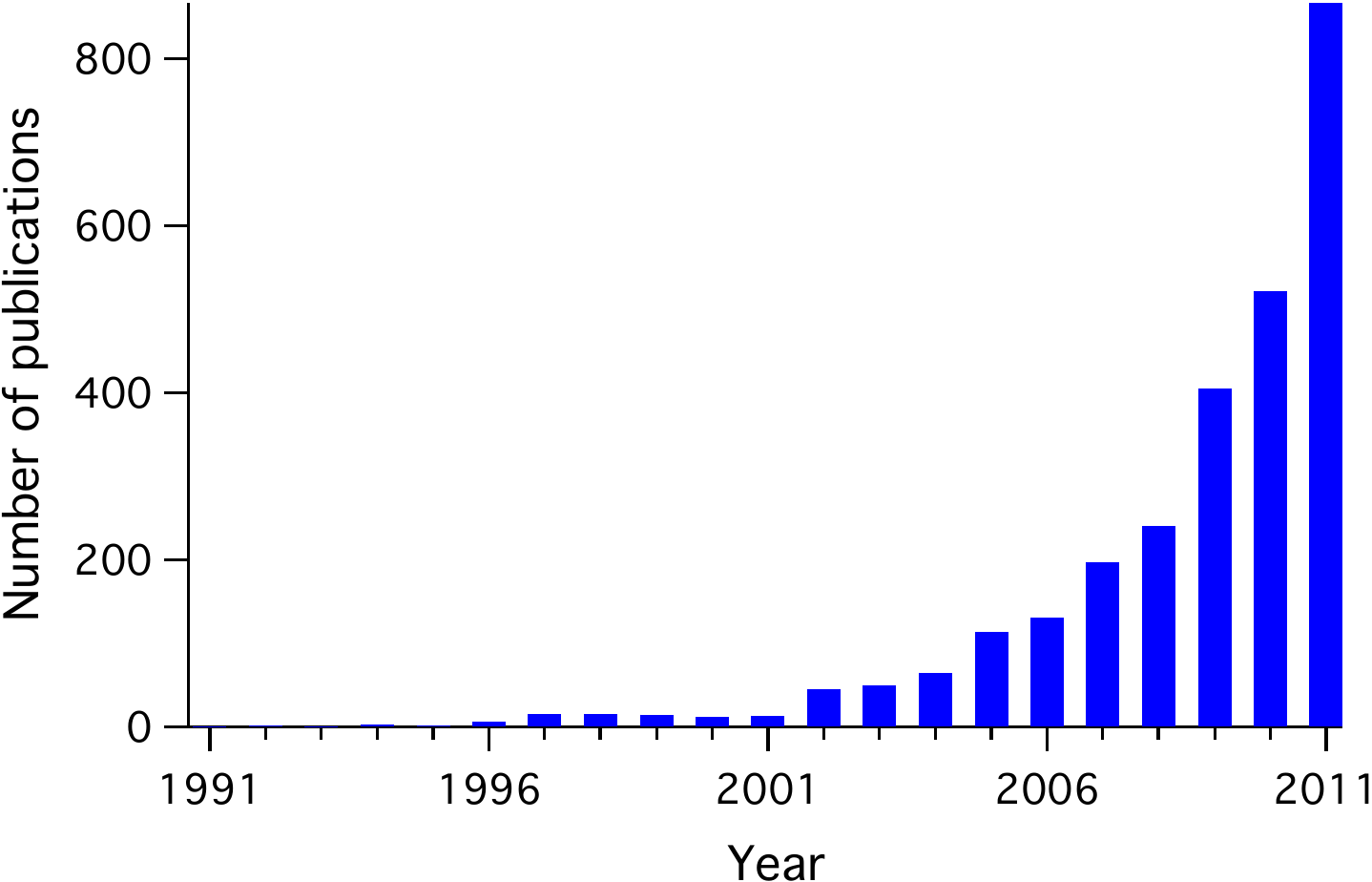}}
\caption[vdw_citations]
{The number of dispersion corrected DFT studies has increased considerably in recent years.
The total number of studies performed is difficult to establish precisely, however, an estimate
is made here by illustrating  the number of papers that cite at least one of 16 seminal works in the 
field (Refs. \onlinecite{rapcewicz1991,andersson1996,dobson1996,elstner2001,wu2001,wu2002,dion2004,lilienfeld2004,grimme2004,becke2005real,becke2005,johnson2005,becke2005df,grimme2006d2,tkatchenko2009,grimme2010}).
(Data from Web of Knowledge, July 2012 for the years 1991-2011)
}
\label{fig_vdw_citations}
\end{figure}

Although DFT is in principle exact, in practice approximations must be made for how electrons interact with each other. 
These interactions are approximated with so-called exchange-correlation (XC) functionals and
much of the success of DFT stems from the fact that XC functionals with very simple forms often yield accurate results.
However there are situations where the approximate form of the XC functional leads to problems. 
One prominent example is the inability of ``standard'' XC 
functionals to describe long-range electron correlations, otherwise known as electron dispersion forces; 
by standard XC functionals we mean the local density approximation (LDA), generalized gradient approximation (GGA) functionals 
or the hybrid XC functionals.
The ``lack'' of dispersion forces -- often colloquially referred to as van der Waals (vdW) forces -- 
is one of the most significant problems with modern DFT
and, as such, the quest for DFT-based methods which accurately account for dispersion
is becoming one of the hottest topics in computational chemistry, physics, and materials science.
Fig.~\ref{fig_vdw_citations} underlines this point, where it can be seen that over 800 dispersion-based DFT
studies were reported 2011 compared to fewer than 80 in the whole of the 1990s.

\begin{figure}[h]
\centerline{
\includegraphics[width=6cm]{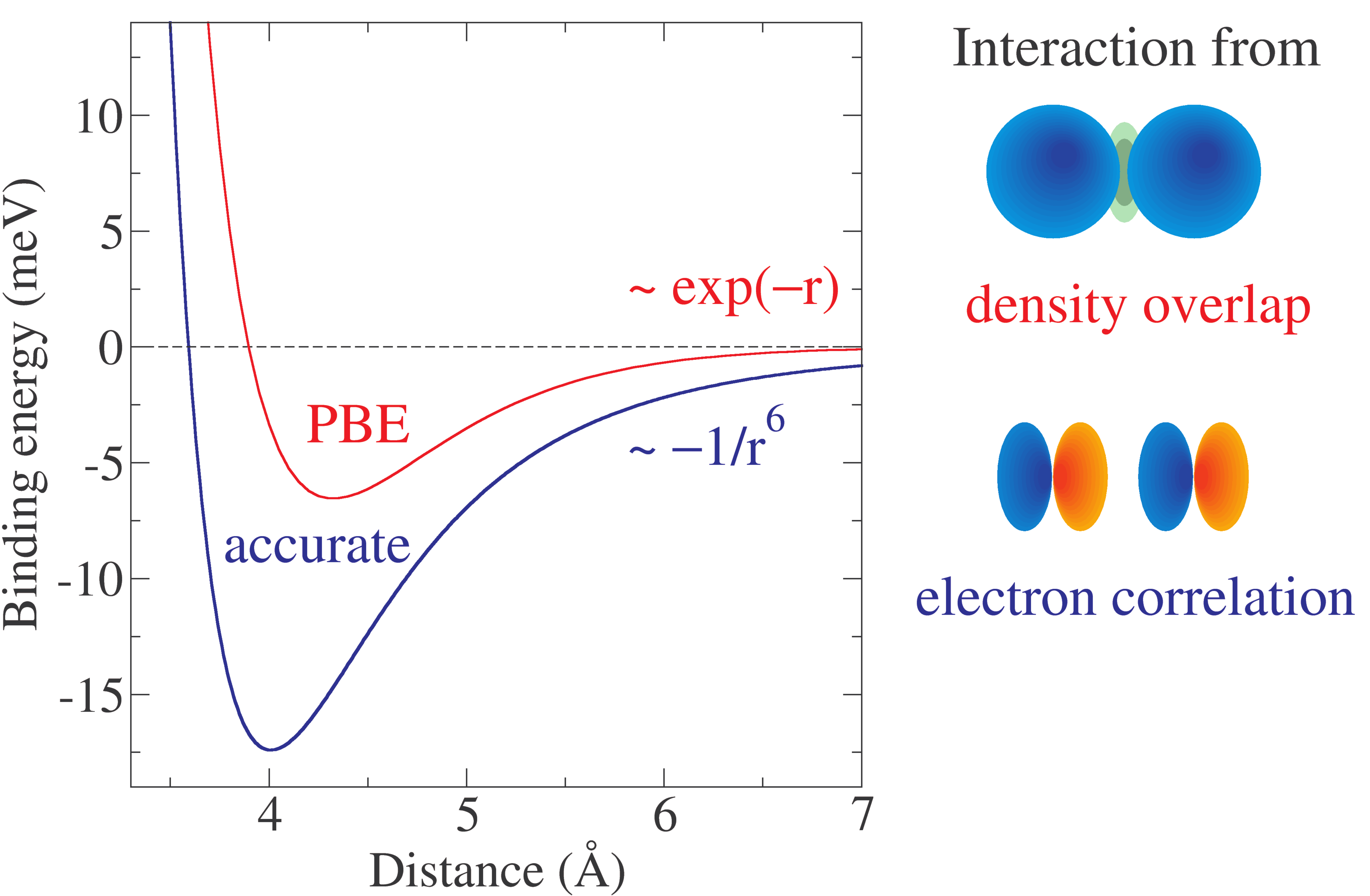}}
\caption[Kr dimer]
{Binding curves for the Kr dimer obtained with the PBE exchange-correlation functional and an accurate model
potential.\cite{tang2003}
Dispersion originates from fluctuations in the electron density which polarize different atoms and molecules, 
as shown schematically in the lower right diagram. 
The interaction then exhibits the well-known $-1/r^6$ decay.
This $-1/r^6$ decay is not reproduced with PBE (or other semi-local functionals) which instead gives an exponential decay 
for the interaction because PBE relies on the overlap of density to obtain the interaction.}
\label{fig_kr_dimer}
\end{figure}

Dispersion can be viewed as an attractive interaction originating from the response of electrons
in one region to instantaneous charge density fluctuations in another. 
The leading term of such an interaction is instantaneous dipole-induced dipole which gives rise to the well known $-1/r^6$ decay of the 
interaction energy with interatomic separation $r$.
Standard XC functionals don't describe dispersion because: (a)
instantaneous density fluctuations are not considered; and (b)
they are ``short-sighted" in that they consider only local properties to calculate
the XC energy.
The consequence for two noble gas atoms, for example, is that these functionals give binding or repulsion
only when there is an overlap of the electron densities of the two atoms.
Since the overlap decays exponentially with the interatomic separation, so too does any binding.
We show  this in Fig.~\ref{fig_kr_dimer} for one of the most widely used GGAs -- the so-called Perdew-Burke-Ernzerhof 
(PBE) functional\cite{perdew1996} -- for a binding curve between two Kr atoms.

The binding of noble gases is a textbook dispersion bonded system. 
However, it has become increasingly apparent that dispersion can contribute significantly to the binding
of many other types of materials, such as biomolecules, adsorbates, liquids, and solids.
Fig.~3 illustrates a more ``real world'' example, where an accurate description of 
dispersion is critical. 
The figure reports binding energies obtained from PBE and an accurate reference method for 
DNA base pairs in two different configurations.
In the first configuration the binding between the base pairs is dominated 
by hydrogen bonding. 
Hydrogen bonding is governed mainly by electrostatics and a standard functional such as 
PBE predicts reasonable hydrogen bond strengths and as a result the stability of the dimer
is quite close (within 15\%) to the reference value.\cite{jurecka2006} 
However, in the other ``stacked" configuration the binding is dominated 
by dispersion forces and for this isomer PBE is hopeless, hardly predicting any binding between the base 
pairs at all. 
This huge variability in performance is far from ideal and since the stacked arrangement of base pairs 
is a common structural motif in DNA, the result suggests that
DNA simulated with PBE would not be stable.

\begin{figure}[h]
\centerline{
\includegraphics[width=6.0cm]{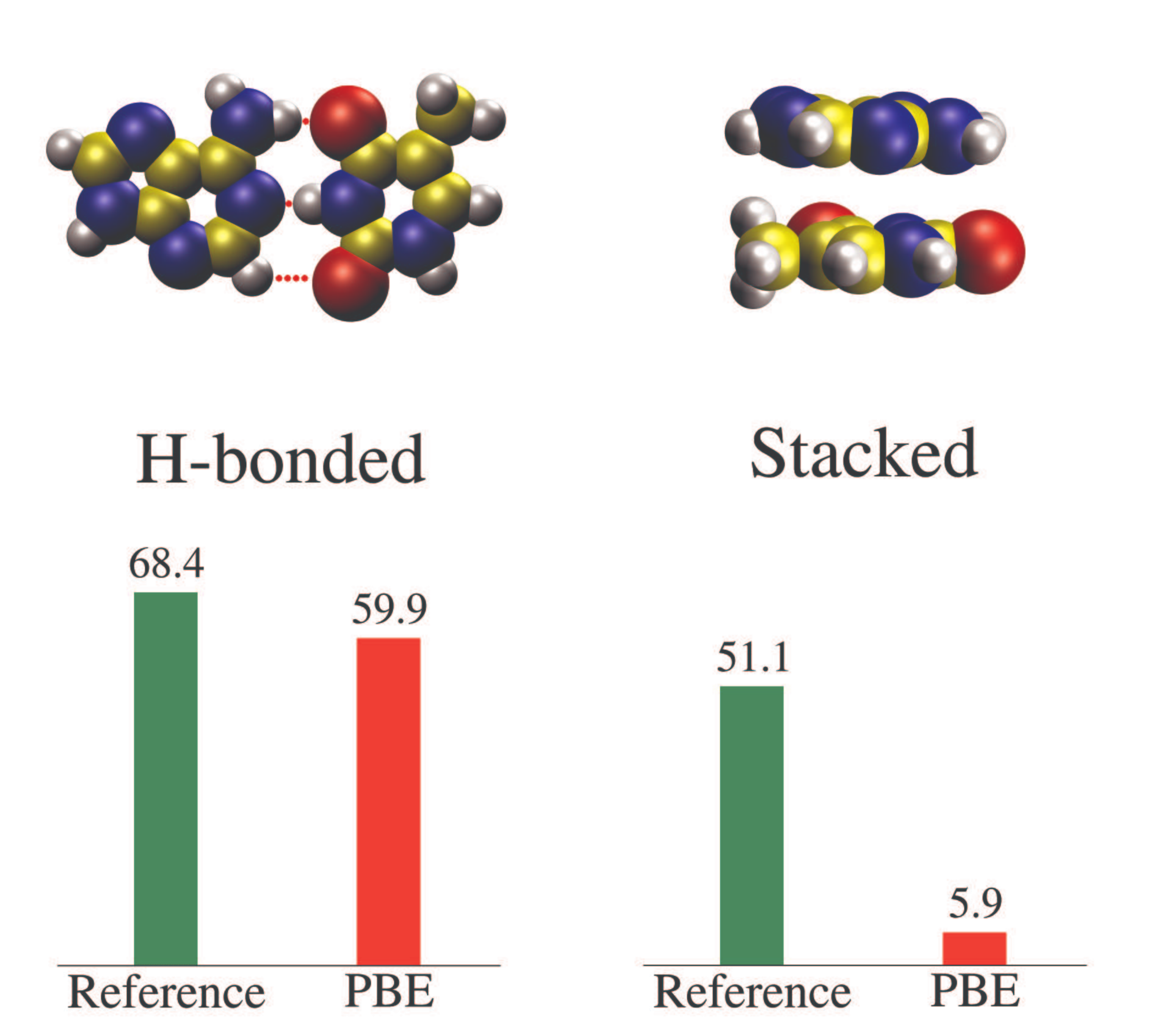}}
\caption[AT dimer]
{Two binding configurations of the DNA base pairs adenine and thymine. A hydrogen bonded structure is shown
on the left (hydrogen bonds indicated by red dots) and a ``stacked" geometry on the right.
For the hydrogen bonded configuration 
a standard XC functional such as PBE gives a binding energy that is in rather good agreement with the reference value
(lower part of the graph, data in kJ/mol).\cite{jurecka2006}
In contrast, PBE fails to describe the binding of the ``stacked" configuration where dispersion interactions contribute
significantly to the intermolecular binding.
}
\label{fig_AT}
\end{figure}

As we will see throughout this article there are many other examples of the importance of 
dispersion to the bonding of materials and in recent years a 
plethora of schemes has been proposed to treat dispersion within DFT. 
Here we will discuss some of the main approaches developed and in the process attempt to provide a useful classification of them.
We also highlight some of the obstacles that
must be overcome before improved DFT-based methods for including dispersion are available.
By its nature, this article is a limited personal view that cannot cover every development in this thriving field. 
For the most part we have tried to keep the overview simple, and have aimed it primarily at newcomers to the field,
although we do go into more depth at the later stages. 
For more detailed discussions of some of the methods shown here the interested reader should consult the reviews 
of Grimme,\cite{grimme2011rev} Tkatchenko~{\it et al.},\cite{tkatchenko2010rev} Johnson~{\it et al.}\cite{johnson2009rev} or
others.\cite{burns2011,grafenstein2009,vydrov2011chap,eshuis2012rev}
Some relevant developments are also discussed in the reviews of Sherrill\cite{sherrill2010} and 
Riley~{\it et al.}\cite{riley2010} which are more focused on post Hartree-Fock (HF) methods.

\section{A classification of the common DFT-based dispersion methods}

Many DFT-based dispersion techniques have been developed and rather than simply listing them all, it is useful to try to classify them.
A natural way to do this is to consider the level of approximation each method makes in obtaining the long range 
dispersion interactions, that is, the interactions between well separated fragments where dispersion
is clearly defined.
In doing this it turns out that groups of methods which exploit similar approximations emerge.  
Here, we simply identify these groupings and rank them from the most approximate to the more sophisticated approaches in a manner that 
loosely resembles the well-known ``Jacob's ladder" of functionals introduced by Perdew.\cite{perdew2001ladder}
Therefore, by analogy, we introduce a ``stairway to heaven" for long range dispersion interactions and 
place each group of dispersion correction schemes on a different step of the stairway.
In complete analogy to the ladder, when climbing the stairway progressively higher overall accuracy can be
expected until exact results, and thus heaven, are reached. 
We stress the point ``overall accuracy'' because, as with the ladder, climbing the stairway does not necessarily mean 
higher accuracy for every particular problem but rather a smaller probability to fail, i.e., a statistical 
improvement in performance.\cite{krieg2010} 
A schematic illustration of the stairway is shown in Fig. 4 and in the following sections each step on the stairway is
discussed.

\begin{figure}[h]
\centerline{
\includegraphics[height=6cm]{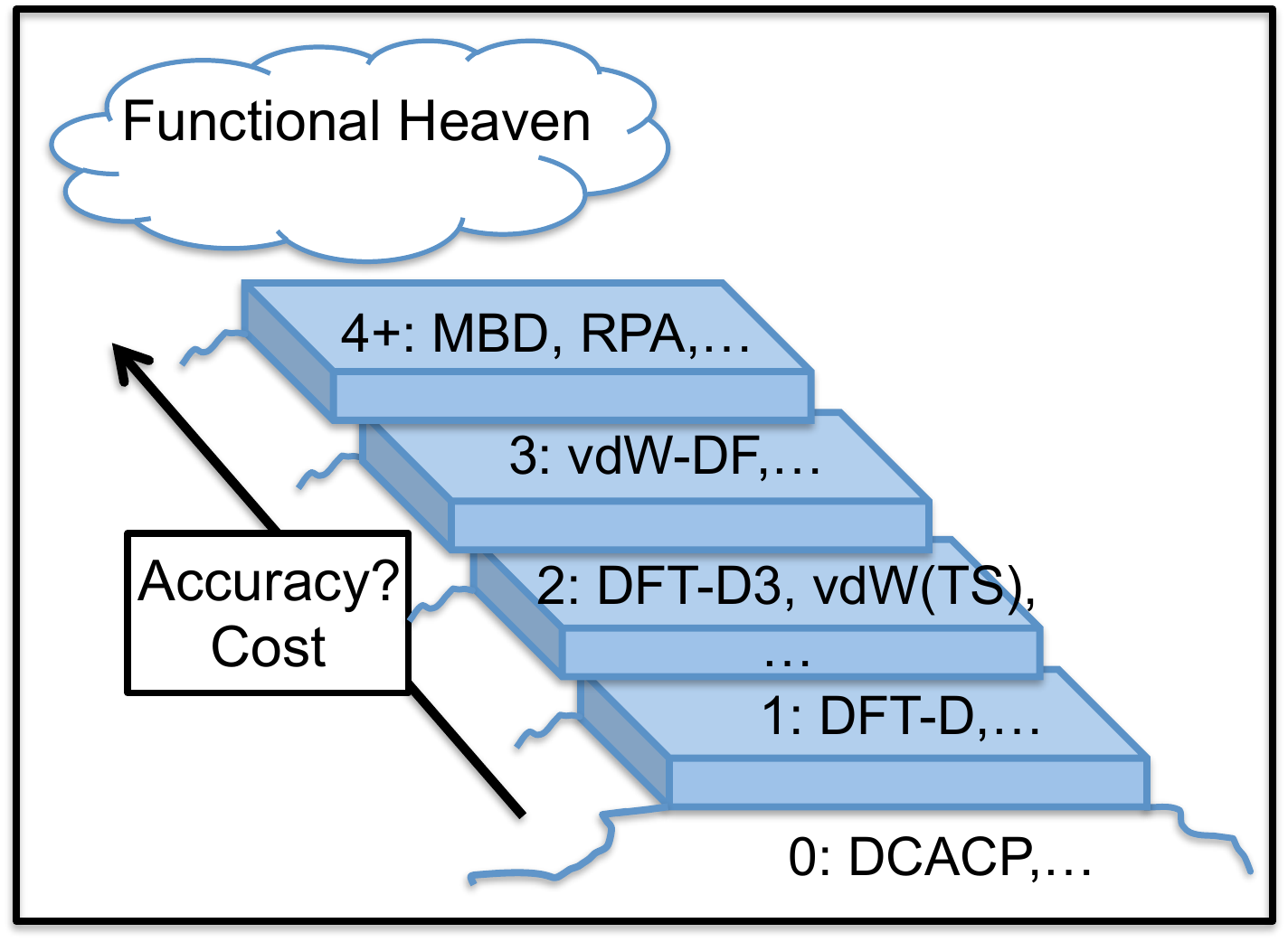}}
\caption[Three things]
{In analogy with the Jacob's ladder classification of functionals the ``stairway to heaven" is used here
to classify and group DFT-based dispersion correction schemes. 
At ground level are methods
which don't describe the long range asymptotics. Simple $C_6$ correction schemes sit on
the first step, on the second step are approaches that utilise environment dependent $C_6$ corrections. 
The long range density functionals sit on step 3 and on step 4 and above are approaches which
go beyond pairwise additive determinations of dispersion.
Upon climbing each step of the stairway the
level of approximation is reduced and the overall accuracy is expected to increase.
}
\label{fig_lad_stair}
\end{figure}

%
\subsection{Ground -- Binding with incorrect asymptotics}
%

First, the ground is occupied by methods that simply {\sl don't} describe the long range asymptotics. 
These approaches give incorrect shapes of binding curves and underestimate the binding of well separated
molecules. 
Some of these methods are, nevertheless, being used for weakly bonded systems.
Although this might seem odd, this somewhat misguided approach is used because some standard DFT functionals 
bind dispersion bonded systems at short separations.
A prominent example is the LDA
which has been used to study systems where dispersion plays a major role such as graphite or noble gases on metals.
However, the results with LDA for dispersion bonded systems have limited and inconsistent accuracy and the asymptotic form of
the interaction is incorrect.
More promising approaches on the ground level of our stairway are density functionals specifically fitted to reproduce 
weak interactions around minima as well as specially adapted pseudopotentials.

The ``Minnesota functionals"\cite{zhao2007} are an example of a new breed of functionals that are fitted 
to a dataset that includes binding energies of dispersion bonded dimers, amongst other properties.
Although these functionals can describe binding accurately at separations around minima, they suffer
from the same incorrect asymptotics as the LDA does.
On the plus side, the reference data used in their construction also contains properties other than weak interactions
(e.g. reaction barriers), so that they can be rather accurate for general chemistry problems.
This is a clear advantage over ``proper" dispersion correction schemes which often utilise 
GGA functionals that can be less accurate for such problems.

For electronic structure codes that make use of pseudopotentials, dispersion 
can also be modelled by adding a specially constructed pseudopotential projector.
Within this class are the dispersion corrected atom-centered potentials 
(DCACP)\cite{lilienfeld2004} and the local atomic potentials (LAP) methods.\cite{sun2008LAP}
These approaches have shown a lot of promise,\cite{tapavicza2007}
however, effort is required to carefully fit the potentials for each element and they are not easily transferable to all-electron methods. 

\subsection{Step one -- Simple $C_6$ corrections}

The basic requirement for any DFT-based dispersion scheme should be that it 
yields reasonable $-1/r^6$  asymptotic behavior
for the interaction of particles in the gas phase, where $r$ is the distance between the particles.
A simple approach for achieving this is to add an additional energy term
which accounts for the missing long range attraction.
The total energy then reads
\begin{equation}
E_{\rm tot}=E_{\rm DFT}+E_{\rm disp}\,,
\end{equation}
where $E_{\rm DFT}$ is the DFT total energy computed with a given XC functional. 
The dispersion interaction is given by
\begin{equation}
E_{\rm disp}=-\sum\limits_{\rm A, B} C^{\rm AB}_6/r^6_{\rm AB}\,,
\label{eq_step1a}
\end{equation}
where the dispersion coefficients $C^{\rm AB}_6$ depend on the elemental pairs A and B. 
Within this approach dispersion is assumed to be pairwise additive and can therefore be calculated as a sum over all pairs of atoms A and B.
Methods on step 1 of the stairway use coefficients that are tabulated, isotropic (i.e. direction independent) and constant,
and these methods are generally termed ``DFT-D".

Mostly because of the simplicity and low computational cost this pairwise $C_6/r^6$ correction scheme is widely used.
%
Nonetheless, it has at least four clear shortcomings which limit the accuracy one can achieve with it.
First, the $C_6/r^6$ dependence represents only the leading term of the correction and neglects both 
many-body dispersion effects\cite{dobson2006} and faster decaying terms such as the $C_8/r^8$ or $C_{10}/r^{10}$ interactions.
%
Second, it is not clear where one should
obtain the $C_6$ coefficients. Various formulae, often involving experimental input (ionization potentials and polarizabilities), 
have been proposed for this.\cite{wu2001,elstner2001,wu2002,grimme2004}
However, this reliance on experimental data limited the set of elements that could be treated 
to those present in typical organic molecules.
The third issue is that the $C_6$ coefficients are kept constant during the calculation, and so effects of 
different chemical states of the atom or the influence of its environment are neglected.
The fourth drawback, which we discuss later, is that the
$C_6/r^6$ function diverges for small separations (small $r$) and this divergence must be removed.

For a more widely applicable method a more consistent means of deriving the dispersion coefficients is required. 
In 2006 Grimme published one such scheme, referred to as DFT-D2.\cite{grimme2006d2} 
In this approach the dispersion coefficients are calculated from a formula which couples ionization potentials 
and static polarizabilities of isolated atoms.
Data for all elements up to Xe are available and this scheme is probably the most widely used 
method for accounting for dispersion in DFT at present.
Whilst incredibly useful, DFT-D2 is also not free from problems. 
In particular, for some elements 
arbitrary choices of dispersion coefficients still had to be made. For example, alkali and alkali earth atoms
use coefficients that are averages of the previous noble gas and group III atom.
Furthermore, the dispersion energy, $E_{\rm disp}$, is scaled according to the XC functional used and as a result
the interaction energy of two well separated monomers is not constant but sensitive to the choice of XC functional.

\begin{figure}[h]
\centerline{
\includegraphics[height=6cm]{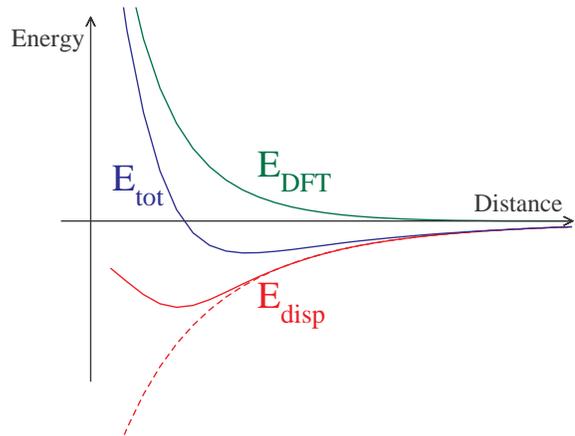}}
\caption[Medium range]
{Schematic illustration of a binding curve ($E_{\rm tot}$) obtained from a dispersion corrected DFT calculation and the contributions to it from the regular DFT energy ($E_{\rm DFT}$) and the dispersion correction ($E_{\rm disp}$). The function $-1/r^6$ used to model the dispersion correction diverges for small $r$ (red dashed curve) and must be damped (solid red curve).
Since the details of the damping strongly affect the position of the energy minima on the binding curve, 
the damping function needs to be fit to reference data. In addition, the damping function will be different for different XC functionals used to obtain $E_{\rm DFT}$.
}
\label{fig_medium}
\end{figure}

With the simple $C_6/r^6$ correction schemes the dispersion correction 
diverges at short inter-atomic separations and so must be ``damped''.
The damped dispersion correction is typically given by a formula like
\begin{equation}
E_{\rm disp}=-\sum\limits_{\rm A, B} f(r_{\rm AB}, {\rm A}, {\rm B})C^{\rm AB}_6/r^6_{\rm AB}\,,
\label{eq_step1}
\end{equation}
where the damping function $f(r_{\rm AB}, {\rm A}, {\rm B})$ is equal to one for large $r$
and decreases $E_{\rm disp}$ to zero or to a constant for small $r$.
We illustrate the divergence and a possible damping by the red curves in Fig.~\ref{fig_medium}.
How the damping is performed is a thorny issue because the shape of the underlying binding 
curve is sensitive to the XC functional used and so the 
damping functions must be adjusted so as to be compatible with each exchange-correlation or exchange functional.
%
This fitting is also sensitive to the definition of atomic size (van der Waals radii are usually used) and must be done carefully since 
the damping function can actually affect the binding
energies even more than the asymptotic $C_6$ coefficients.\cite{hanke2011}
The fitting also effectively includes the effects of 
$C_8/r^8$ or $C_{10}/r^{10}$ and higher contributions, although 
some methods treat them explicitly.\cite{becke2007,grimme2010}
An interesting damping function has been proposed within the so-called DFT coupled cluster approach (DFT/CC), 
it has a general form that can actually force the dispersion correction to be repulsive.\cite{bludsky2008}

Before leaving the simple $C_6/r^6$ schemes we note that although the approaches developed 
by Grimme are widely used, other parameterizations, which differ in, e.g., the combination 
rules used, are available.\cite{zimmerli2004,neumann2005,jurecka2007jcc,bludsky2008}
Furthermore, functionals such as B97-D\cite{grimme2006d2} and $\omega$B97X-D\cite{chai2008} have been specifically designed 
to be compatible with this level of dispersion correction.

\subsection{Step two -- Environment-dependent $C_6$ corrections}

A problem with the simple ``DFT-D" schemes is that the dispersion 
coefficients are predetermined and constant quantities.
Therefore the same coefficient will be assigned to an element no matter what its oxidation or hybridization
state. 
However the errors introduced by this approximation can be large, e.g. the 
carbon $C_6$ coefficients can
differ by almost 35\% between the $sp$ and $sp^3$ hybridized states.\cite{wu2002}
Thus the emergence of methods where the $C_6$ coefficients vary with 
the environment of the atom has been a very welcome development.
We put these methods on the second step of our stairway.
They still use Equation~\ref{eq_step1} to obtain the dispersion correction
and, as with the step~1 methods, some reference data (such as atomic polarizabilities)  
is used to obtain the $C_6$ coefficients.

We now discuss three step 2 methods: DFT-D3 of Grimme, the approach of Tkatchenko and Scheffler (vdW(TS)), 
and the Becke-Johnson model (BJ).
The unifying concept of these methods is that the dispersion coefficient of an atom in a molecule depends on the effective volume of the atom. 
When the atom is ``squeezed", its electron cloud becomes less polarizable leading to a decrease of the $C_6$ coefficients.

Grimme {\it et al.}~\cite{grimme2010} capture the environmental dependence of the $C_6$ coefficients 
by considering the number of neighbors each atom has.
When an atom has more neighbors it is thought of as getting squeezed and as a result the $C_6$ coefficient decreases.
This effect is accounted for by having a range of precalculated $C_6$ coefficients for various pairs of elements 
in different reference (hybridization) states. 
In the calculation of the full system the appropriate $C_6$ coefficient is assigned to 
each pair of atoms according to the current number of neighbors.
The function calculating the number of neighbors is defined in such a way that it 
continuously interpolates between the precalculated reference values.
Therefore if the hybridization state of an atom changes during a simulation, the $C_6$ coefficient can also change continuously.
Despite the simplicity of this approach, termed ``DFT-D3", the dispersion coefficients it produces are pretty accurate. 
Specifically, based on the reference data of Meath and coworkers there is a mean absolute percentage deviation
(MAPD) in the $C_6$ coefficients of $8.4\%$. 
Moreover, the additional computational cost is negligible since the number of neighbors can be obtained quickly.

In 2009 Tkatchenko and Scheffler\cite{tkatchenko2009} proposed a method which 
relies on reference atomic polarizabilities and reference atomic $C_6$ coefficients\cite{chu2004} to calculate the dispersion energy.
These quantities are sufficient to obtain the $C_6$ coefficient for a pair of unlike atoms.\cite{tang1969}
To obtain environment dependent dispersion coefficients effective atomic volumes are used.
During the calculation on the system of interest the electron density of a molecule is divided between the individual 
atoms (the Hirshfeld partitioning scheme is used)
and for each atom its corresponding density is compared to the density of a free atom.
This factor is then used to scale the $C_6$ coefficient of a reference atom which changes the value of the dispersion energy.
The accuracy of the final (isotropic, averaged) $C_6$ coefficients is quite high, with the 
MAPD on the data of Meath and coworkers being only $5.4\%$.
However, it is not yet clear if the scaling of $C_6$ coefficients with volume will be accurate for more
challenging cases such as different oxidation states in e.g. ionic materials.

The most complex step 2 method is the BJ model,\cite{becke2005real,becke2005,johnson2005,becke2005df,becke2007} which exploits the fact that around an 
electron at $r_1$ there will be a region of electron density depletion, the so-called
XC hole. 
Even for symmetric atoms, this creates asymmetric electron density and thus
non-zero dipole and higher-order electrostatic moments, 
which causes polarization in other atoms to an extent given by their polarizability.
The result of this is an attractive interaction with the leading term being dipole-induced dipole.
The biggest question in the BJ model is how to quantify the XC hole and in the method 
it is approximated as the exchange-only hole (calculated using the Kohn-Sham orbitals).
An average over the positions $r_1$ of the reference electron then gives the average square of the hole dipole moment, denoted $\langle d_{\rm x}^2 \rangle$
which together with the polarizabilities $\alpha$ enters the formula for the dispersion coefficient
\begin{equation}
C_6^{\rm AB}={\alpha_{\rm A} \alpha_{\rm B} \langle d_{\rm x}^2 \rangle_{\rm A} \langle d_{\rm x}^2 \rangle_{\rm B} \over 
\langle d_{\rm x}^2 \rangle_{\rm A} \alpha_{\rm B}+\langle d_{\rm x}^2 \rangle_{\rm B} \alpha_{\rm A}}\,.
\end{equation}
This is, in fact, very similar to the vdW(TS) formula but in vdW(TS) precalculated $C_6$ coefficients are used instead of the 
hole dipole moments.
The BJ model is very intriguing and several authors have studied how it relates to formulae derived from perturbation 
theory.\cite{angyan2007BJ,ayers2009BJ,hesselmann2009BJ}

In the BJ model the $C_6$ coefficients are altered through two effects.
First, the polarizabilities of atoms in molecules are scaled from their reference atom values according to their effective atomic volumes.
Second, the dipole moments respond to the chemical environment through changes of the exchange hole, although this 
effect seems to be difficult to quantify precisely.
The details of how to obtain atomic volumes and the exchange hole are known to affect the results obtained 
to some extent,\cite{becke1989,kannemann2010,steinmann2011} 
but overall the accuracy of the asymptotic $C_6$ coefficients obtained from the BJ model is quite satisfactory, with a
MAPD of 12.2~\% for the data of Meath and coworkers.\cite{grimme2010}
Compared to the two previous step 2 approaches, however, the computational cost is relatively high, 
in the same ballpark as the cost of a hybrid functional.\cite{burns2011}

\subsection{Step three -- Long-range density functionals}

The methods discussed so far require predetermined input parameters to calculate the dispersion
interaction, either the $C_6$ coefficients directly or the atomic polarizabilities.
Now we discuss approaches that do not rely on external input parameters 
but rather obtain the dispersion interaction directly from the electron density.
This, in principle, is a more general strategy and thus we place these methods on step 3 of our stairway.
The methods have been termed non-local correlation functionals since they add non-local 
(i.e. long range) correlations to local or semi-local correlation functionals. 

The non-local correlation energy $E_{\rm c}^{\rm nl}$ is calculated from
\begin{equation}
E_{\rm c}^{\rm nl}=\iint {\d}r_1 {\d}r_2 n(r_1) \varphi(r_1,r_2) n(r_2)\,.
\label{eq_vdw_nl}
\end{equation}
This is a double space integral where $n(r)$ is the electron density and $\varphi(r_1,r_2)$ is some integration kernel.
%
The kernel is analogous to the classical Coulomb interaction kernel $1/|r_1-r_2|$ but with a more complicated formula used 
%
for $\varphi(r_1,r_2)$ with $\mathcal{O}(1/|r_1-r_2|^6)$ asymptotic behavior.
The formula has a pairwise form and thus ignores the medium between points $r_1$ and $r_2$.
Various forms for $E_{\rm c}^{\rm nl}$ were proposed in the nineties~\cite{rapcewicz1991,dobson1996,andersson1996} but 
were restricted to non-overlapping fragments.
This rather severe limitation was removed by Dion~{\it et al.} in 2004~\cite{dion2004} who proposed a functional form
which can be evaluated for overlapping molecules and for arbitrary geometries.
Within this approach the XC energy $E_{\rm xc}$ is calculated as
\begin{equation}
E_{\rm xc}=E_{\rm x}^{\rm GGA}+E_{\rm c}^{\rm LDA}+E_{\rm c}^{\rm nl},
\label{eq_vdw_xc}
\end{equation}
where the terms on the right hand side are the exchange energy in the revPBE approximation,\cite{zhang1998} the LDA correlation energy, 
and the non-local correlation energy term, respectively.
This method has been termed the van der Waals density functional (vdW-DF).
vdW-DF is a very important conceptual development since it adds a description of dispersion directly within a DFT functional
and combines correlations of all ranges in a single formula.

Since the development of the original vdW-DF a number of follow up studies have aimed at understanding and improving 
the performance of the method. 
First, it turns out that vdW-DF tends to overestimate the long range dispersion interactions:  
Vydrov and van Voorhis have shown that when the $C_6$ coefficients are evaluated the errors can be as large as $\sim$30\%.\cite{vydrov2009prl}
To address this these authors proposed (computationally cheaper) functionals that reduce the 
average errors by approximately 50\%.\cite{vydrov2009jcp,vydrov2009prl,vydrov2010pra,vydrov2010jcp}
The developers of the original vdW-DF tried to address its tendency to overbind at large separations by proposing vdW-DF2.\cite{murray2009,lee2010} 
This functional, which involves changes to both the exchange and non-local correlation components
tends to improve 
the description of the binding around energy minima, however, the $C_6$ coefficients it predicts are considerably underestimated.\cite{vydrov2010pra}
Second, aside from the particular form of $E_{\rm c}^{\rm nl}$, the choice of the exchange functional used in Equation~\ref{eq_vdw_xc}
has received considerable attention.
The original revPBE exchange functional chosen for vdW-DF often leads to too large intermolecular binding distances
and inaccurate binding energies.
To remedy these problems alternative ``less repulsive" exchange functionals have been proposed\cite{klimes2010,klimes2011,cooper2010,wellendorff2012} which
when incorporated within the vdW-DF scheme (Equation~\ref{eq_vdw_xc}) lead to much improved accuracy. 
Of these the ``optB88" and ``optPBE" exchange functionals have been shown to offer very good performance for a wide 
range of systems.\cite{klimes2010,carrasco2011,klimes2011,mittendorfer2011}

Initially the non-local correlation functionals came, to a lesser or greater extent,\cite{vydrov2008,gulans2009} 
with a higher computational cost than GGAs or hybrid functionals.
However, thanks to the work of Rom\'{a}n-P\'{e}rez and Soler the computational cost of vdW-DF is now comparable to 
that of a GGA.\cite{soler2009}
In addition, self-consistent versions of vdW-DF and several of its offspring are now implemented in widely distributed
DFT codes such as Siesta,\cite{soler2009} VASP,\cite{klimes2011} QuantumESPRESSO,\cite{kolb2011} and QChem.\cite{vydrov2008}
Other ways to reduce the cost of vdW-DF calculations have also been reported, for example, Silvestrelli has shown\cite{silvestrelli2008,silvestrelli2009surf}
that utilizing Wannier functions to represent the electron density allows for an analytic evaluation 
of the functionals proposed in the nineties.\cite{dobson1996,andersson1996} 
Particularly interesting is the local response dispersion (LRD) approach of Sato and Nakai\cite{sato2009,sato2010} which
yields very accurate $C_6$ coefficients. 
Overall, by not relying on external reference data, the step 3 approaches
are less prone to fail for systems outside of the reference or fitting space of step 2 methods.
However, very precise calculations of the dispersion energy are difficult and the formulae underlying vdW-DF and similar
approaches can be somewhat complicated.

\subsection{Higher steps -- Beyond pairwise additivity}

The main characteristic of the methods discussed so far is that they consider dispersion to be pairwise additive. 
As a consequence the interaction energy of two atoms or molecules remains constant 
no matter what material separates them and all atoms interact ``on their own" with no consideration 
made for collective excitations.\cite{misquitta2010}
%
While such effects don't seem to be crucially important in the gas phase, especially for small molecules, 
they are important for adsorption or condensed matter systems where the bare interaction is screened.\cite{zhang2011}
We now briefly discuss some of the methods being developed which treat dispersion beyond the pairwise approximation.
The range of approaches is quite wide, from methods based on atom centered interactions, to methods involving
density, to methods using electron orbitals.
Because of the freshness of most of the methods we discuss them together.

\begin{figure}[h]
\centerline{
\includegraphics[height=4cm]{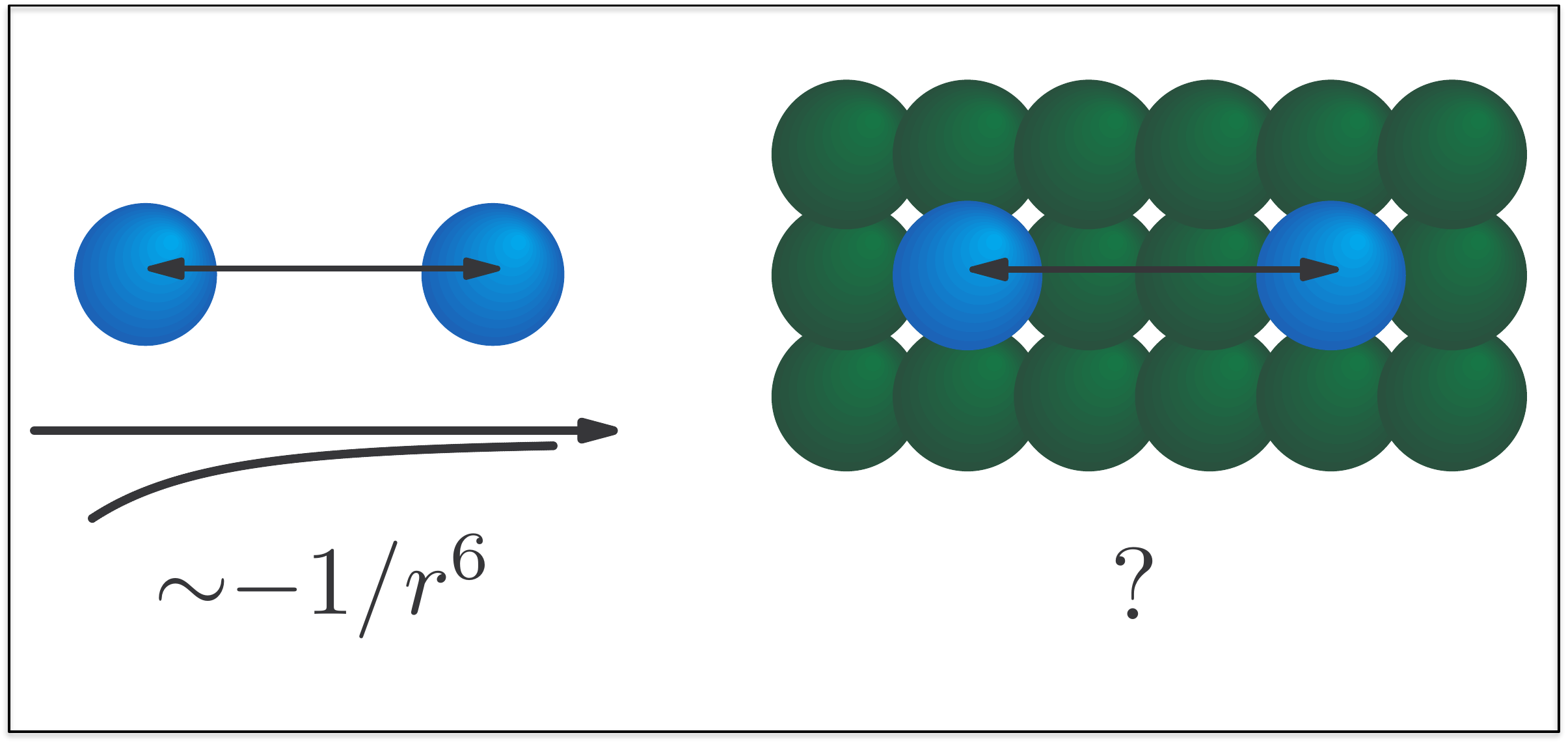}}
\caption[Many body]
{The long-range electron correlations of two isolated atoms can be described by an effective $-1/r^6$ formula.
In a condensed system the interaction is modified (screened) by the presence of the other electrons.
}
\label{fig_manybody}
\end{figure}

Recently, 
the Axilrod-Teller-Muto\cite{axilrod1943,muto1943} formula has been used to extend the atom-centred pairwise approaches 
to include three-body interactions.\cite{grimme2010,lilienfeld2010}
The triple-dipole interaction between three atoms A, B, and C can be written as
\begin{equation}
E^{\rm ABC}=C_{9}^{\rm ABC}{3 \cos(\alpha) \cos(\beta) \cos(\gamma) +1\over r^3_{\rm AB} r^3_{\rm BC} r_{\rm AC}^3}\,,
\end{equation}
where $\alpha$, $\beta$, and $\gamma$ are the internal angles of the triangle formed by the atoms A, B, and C. 
The dispersion coefficient $C_9^{\rm ABC}$ is again obtained from reference data.
Notably, von Lilienfeld and Tkatchenko estimated the magnitude of the three body terms to be $\sim-$25\% (destabilising) of the 
two body term for two graphene layers.\cite{lilienfeld2010}
Since the three-body interaction is repulsive for atoms forming an acute angled triangle, the interaction was found to be
destabilising for stacked configurations of molecular clusters such as benzene dimers. 

The atom-centered approach can be used to approximate the many-body dispersion interaction using a model 
of coupled dipoles.\cite{cole2009,tkatchenko2011}
Here quantum harmonic oscillators with characteristic frequencies occupy each atomic position
and the dispersion interaction is obtained from the shifts of the frequencies of the oscillators upon switching on their interaction.
The model has been applied outside DFT for some time\cite{renne1967,lucas1967,delanaye1978,schmeits1978,sommerfeld2008}
and has recently been used to show the non-additivity of dispersion in anisotropic materials.\cite{kim2006}
The initial applications of this approach termed many-body dispersion (MBD) are quite promising and suggest 
that the higher-order dispersion terms can be important even for systems such as solid benzene.\cite{tkatchenko2011} 
However, getting accurate relations between atoms and oscillator models seems to be rather challenging, 
especially when describing both localized and delocalized electrons.

In the context of DFT, orbitals can be used to calculate the correlation energy using the adiabatic-connection fluctuation-dissipation theorem 
(ACFDT).\cite{dobson1996,furche2005} 
The particular approach which has received the most attention recently is the so-called 
random phase approximation (RPA).\cite{footRPA,furche2001}
Results from RPA have been very encouraging and it exhibits, for example, a consistent accuracy for solids\cite{harl2009,harl2010}
and the correct asymptotic description for the expansion of graphite, a feature which the pairwise methods
fail to capture.\cite{dobson2006,lebegue2010}
RPA is analogous to post-HF methods and indeed direct links have been established.\cite{scuseria2008,jansen2010}
Unfortunately, it also shares with the post-HF methods a high computational cost (the cost increases approximately with the fourth power 
of the system size)\cite{lu2009,eshuis2010} and slow convergence with respect to basis set 
size.\cite{harl2008,zhu2010,toulouse2011,eshuis2011,eshuis2012}
Approaches going beyond RPA are also receiving attention, for example the second-order screened exchange\cite{grueneis2009}
or single excitations corrections.\cite{ren2010}

Another method involving orbitals to obtain the correlation energy is to 
combine DFT with post-HF methods in the hope of exploiting the benefits of each approach.
While the post-HF methods can describe long range interactions accurately (albeit expensively) using orbitals,
DFT is efficient for the effective description of the short range part of the interactions. 
So called range separated methods are an example of this approach where long range correlations are
treated by, e.g., second-order perturbation theory, the coupled cluster 
method, or RPA.
Range separated functionals can be very accurate, can account for many-body interactions, 
and can deliver relatively fast convergence of the correlation
energy with respect to basis set size.\cite{angyan2005,goll2005,goll2006,toulouse2009,janesko2009,paier2010}
Another example from this class are the so called double hybrid functionals, which include Fock exchange and a second order perturbation theory
correlation contribution. 
Generally the coefficients for these contributions are fitted to reproduce reference data.\cite{grimme2006,schwabe2007}
Since the original double hybrid functionals were more concerned with
reaction energies and barrier heights than dispersion, 
dispersion interactions were actually underestimated.
Newer functionals, e.g., mPW2PLYP-D, add a dispersion correction in the sense of step~1 or 2 methods, however,
they are computationally very expensive and at present
prohibitively expensive for most condensed phase and surface studies.\cite{schwabe2007,krieg2010}

\subsection{Summary}

The higher steps of the stairway with the promising schemes that go beyond the pairwise approximation close our overview and classification
of methods 
for treating dispersion within DFT. 
In Table~\ref{tab_meth} we summarize some of the key aspects of the most relevant schemes, such as what 
properties they use to obtain the dispersion contribution (via, e.g., the $C_6$) and the approximate relative
computational cost.
The computational cost of a method is, of course, a key factor since no matter how accurate it is, it
will not be used if the computational cost is prohibitive.
In this regard the cost of the simple pairwise corrections on step~1 is essentially zero compared 
to the cost of the underlying DFT calculation and these methods can be recommended as a good starting point for 
accounting for dispersion. 
Because of the correct asymptotic behavior (at least for gas phase molecules)\cite{dobson2006} they are preferable to methods from ground. 
Compared to the step 1 methods the DFT-D3 and vdW(TS) schemes don't add a significant computational cost, whereas the vdW-DF
approach increases the computational time by about $50\%$ compared to a GGA calculation.
Since the accuracy of these approaches tends to be higher than that of the step~1 methods, they should be preferred 
over the step~1 corrections.
However, when applied to a particular problem, the accuracy of any method used should be tested or verified against 
experimental or other reference data since any approach can, in principle, fail for a specific system.
The methods on higher steps are mostly in development and currently the ACFDT or range separated functionals 
require a computational cost two to four orders of magnitude higher than a GGA calculation which limits their applicability.

\begin{table*}[h]
\caption{Summary of some of the most widely used methods for capturing dispersion interactions in DFT, ordered
according to the ``Step'' they sit on in the stairway to heaven (Fig. 4). 
Information on the reference used for the C$_6$ coefficients, what the C$_6$ depend on, and the additional 
computational cost of each approach compared to a regular GGA calculation is reported.
}
\label{tab_meth}
\centerline{
\begin{ruledtabular}
\begin{tabular}{lcccccc}
\multirow{2}{*}{Method} & \multirow{2}{*}{Step} &\multirow{2}{*}{Reference for C$_6$} & \multirow{2}{*}{C$_6$ depend on} &  Additional& \multirow{2}{*}{Ref.} \\
  & & & &computational cost 
  \footnote{The BJ model and the double hybrids (labelled ``Dbl. hybrids") are more computationally expensive than the simpler ``DFT-D" methods\cite{burns2011} 
and the calculation of the correlation energy in vdW-DF leads to a $\approx$50\% slow-down 
compared to a GGA calculation when done efficiently.\cite{soler2009}} & \\ 
\hline
Minnesota &0& None & N/A & None & e.g.~\onlinecite{zhao2007}\\
DCACP &0& None & N/A & Small &\onlinecite{lilienfeld2004}\\ 
DFT-D &1& Various & Constant & Small &e.g.~\onlinecite{grimme2006}\\
DFT-D3&2& TDDFT  & Structure & Small &\onlinecite{grimme2010}\\
\multirow{2}{*}{vdW(TS)}& \multirow{2}{*}{2}& Polarizabilities& \multirow{2}{*}{Atomic volume} & \multirow{2}{*}{Small}& \multirow{2}{*}{\onlinecite{tkatchenko2009}}\\
   &   &and atomic  $C_6$& & & \\
BJ&2& Polarizabilities & Atomic volume, X hole & Large&\onlinecite{becke2005}\\
LRD&3& $C_6$ calculated & Density & Small &\onlinecite{sato2009}\\
vdW-DF&3& $C_6$ calculated & Density &  $\approx$50\%  &\onlinecite{dion2004}\\
Dbl. hybrids &4&None or as ``-D"& Orbitals & Large &\onlinecite{grimme2006}\\
\end{tabular}
\end{ruledtabular}}
\end{table*}

\section{General performance}

We have commented in passing on how some of the methods perform, mainly 
with regard to the dispersion coefficients. 
Now we discuss accuracy in more detail by focussing on binding energies for a few
representative gas phase clusters, a molecular crystal, and an adsorption problem.
Before doing so, it is important to emphasise two points. 
First, establishing the accuracy of a method is not as straightforward as it might seem since it must be tested 
against accurate reference
data (e.g. binding energies).
Experimental data can be inaccurate and is rarely directly comparable to theory since 
quantum nuclear and/or thermal effects, which affect the experimental values, are often neglected in simulation studies.
Theoretical reference data, on the other hand, is more appropriate but rather hard to 
come by since accurate reference methods such as post-HF methods (MP2, CC) or quantum Monte Carlo (QMC)
are so computationally expensive that they can generally be applied to only very small systems (tens of atoms).\cite{sherrill2010}
Second, the results obtained with many dispersion based DFT approaches  
are often strongly affected by the fitting procedure used when combining the long range dispersion interaction with 
the underlying exchange or XC functionals.

\subsection{Gas phase clusters}

Because of growing computer power and algorithmic improvements, systems of biological importance
such as DNA and peptides can now be treated with DFT. 
As discussed in the introduction, dispersion plays an important role in stabilising the folded state and thus dispersion corrected
functionals are required.
However, since accurate reference data can only be obtained for systems with tens of atoms, it's just not possible to obtain reference 
data for DNA or a protein directly and 
fragments of the large molecules are used to build test sets of binding energies and geometries instead.
One such test set\cite{molnar2009,grafova2010,rezac2011} is the popular S22 data set of Jure\v{c}ka~{\it et al.}.\cite{jurecka2006}
It is useful since it contains 22 different dimers with a range of weak bonding types and with a wide range of interaction energies.
Results for some of the methods are reported in Table~\ref{tab_s22}; specifically we report 
mean absolute deviations (MAD) and mean absolute percentage deviations (MAPD). 
Included in Table~\ref{tab_s22} are results from LDA and semi-local PBE. These functionals fail to reproduce the correct interaction energies, 
yielding MADs of $\sim$10~kJ/mol and MAPDs in excess of 30\%.
Much improved performance is observed with even the simplest pairwise corrections schemes.
Indeed, on all steps of the stairway above ground there is at least one method with 
a MAD below 1.5~kJ/mol (or 0.4~kcal/mol, $\sim$16~meV) and MAPD close to 5\%.
This is very good agreement with the reference data and a clear indication of the recent improvements made in 
the DFT-based description of dispersion. 
As can be seen, the results obtained on lower steps of the stairway can surpass those from methods on 
higher steps. 
This is not that surprising since some of the methods reported in Table~\ref{tab_s22} were actually developed 
by fitting to the S22 data set itself.

\begin{table}[h]
\caption{Mean absolute deviations and mean absolute percentage deviations of different dispersion based DFT methods 
on the S22 data set of Jure\v{c}ka~{\it et al.}.\cite{jurecka2006,podeszwa2010} 
Results are in kJ/mol for MAD and percent for MAPD. 
Results from second order M{\o}ller-Plesset perturbation theory (MP2) -- a widely used post-HF method -- 
are shown for comparison as are results from the LDA and PBE functionals as examples of methods that don't treat dispersion explicitly.}
\label{tab_s22}
\centerline{
\begin{ruledtabular}
\begin{tabular}{lccc}
Method & MAD & MAPD & Ref.\\
\hline 
{\bf Ground} & & &\\
LDA & 8.66  & 30.5 &\onlinecite{klimes_unp}\\
PBE & 11.19 & 57.6 &\onlinecite{klimes_unp}\\
M06-2X& 2.52 & 11.3 &\onlinecite{zhao2008}\\
LAP& 2.51 & 7.8  &\onlinecite{sun2008LAP}\\
\hline
{\bf Step 1} & &  & \\
B97-D2 & 1.51 & 7.3  &\onlinecite{chai2008}\\
$\omega$B97X-D & 0.76 & 5.6 &\onlinecite{chai2008}\\
\hline
{\bf Step 2} & & & \\
BLYP-D3 & 0.96* & -- &\onlinecite{grimme2010}\\
PBE-vdW(TS)& 1.25** & 9.2** &\onlinecite{tkatchenko2009}\\
rPW86-BJ& 1.50 & 6.1 &\onlinecite{kannemann2010}\\
\hline
{\bf Step 3} & &  & \\
LC-BOP+LRD& 0.86 & 4.6 &\onlinecite{sato2010}\\
vdW-DF& 6.10 & 22.0 &\onlinecite{gulans2009}\\
optB88-vdW& 1.18 & 5.7 &\onlinecite{klimes2010}\\
vdW-DF2& 3.94 & 14.7 &\onlinecite{vydrov2010VV10}\\
VV10& 1.35 & 4.5 &\onlinecite{vydrov2010VV10}\\
\hline
{\bf Higher steps}& & & \\
B2PLYP-D3& 1.21* & -- &\onlinecite{grimme2010}\\
$\omega$B97X-2& 1.17 & 8.8 &\onlinecite{chai2009}\\
PBE-MB& -- & 5.4**  &\onlinecite{tkatchenko2011}\\
RPA & 3.29**    &   --    &\onlinecite{eshuis2012}\\
\hline
{\bf post-HF}& & & \\
MP2&3.58 &19.4  &\onlinecite{podeszwa2010}\\
\end{tabular}
\end{ruledtabular}}
* using the original reference values of Jure\v{c}ka~{\it et al.} (Ref.~\onlinecite{jurecka2006})\hfill\\
** using the values of Takatani~{\it et al.} (Ref.~\onlinecite{takatani2010}) which are very similar to 
those of Podeszwa~{\it et al.} (Ref.~\onlinecite{podeszwa2010})\hfill
\end{table}

\begin{figure}[t]
\centerline{
\includegraphics[height=4cm]{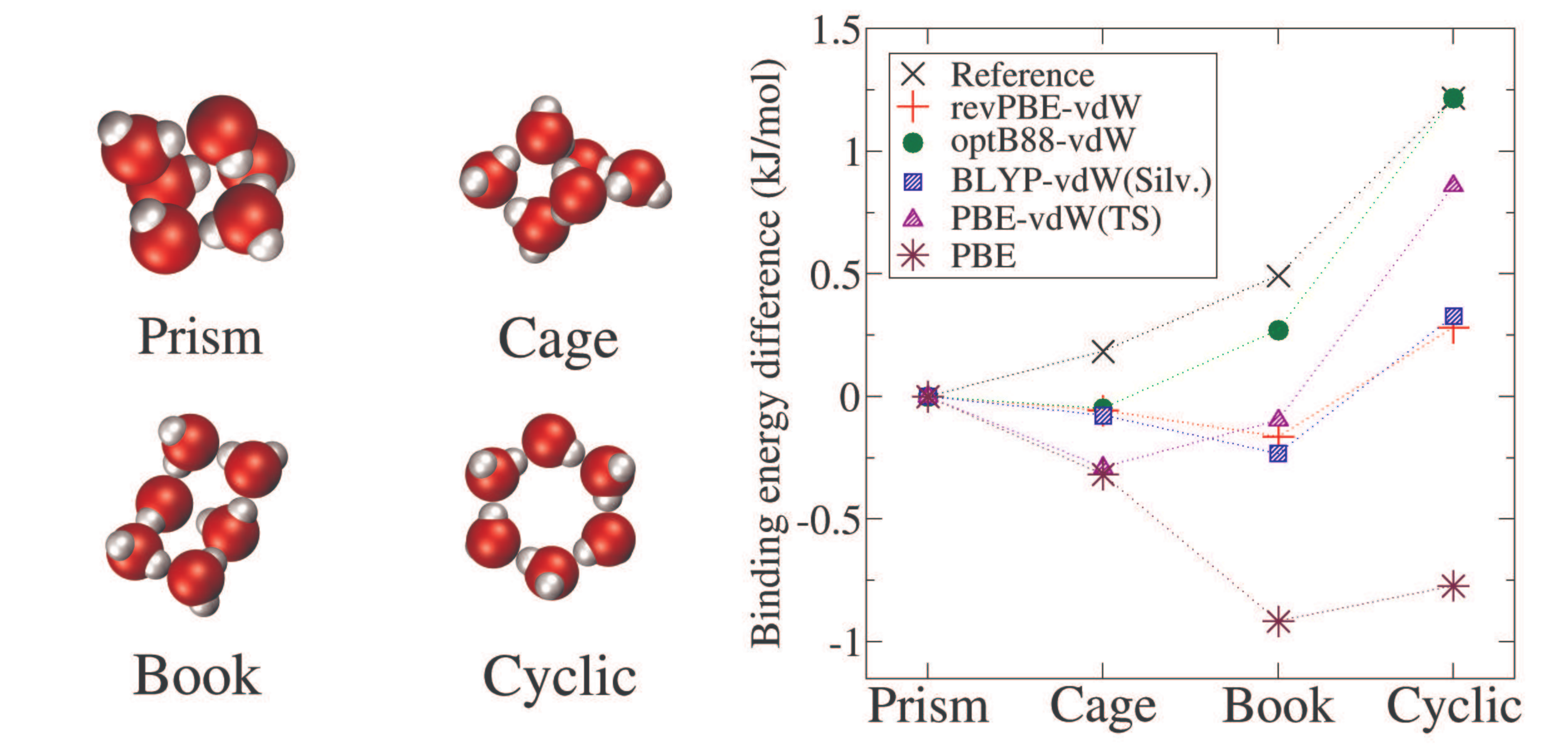}}
\caption[The low energy isomers of water hexamer]
{The four low energy isomers of the water hexamer are an important example where dispersion
interactions play a significant role. In the panel on the right binding energy differences per water molecule
as predicted by different XC functionals are compared to accurate reference data.\cite{santra2008,klimes2010}
While the reference method predicts the ``prism" and ``cage" isomers
to be more stable, standard XC functionals such as PBE give the ``book" and ``cyclic" clusters as the ones with lower energy.
It was shown in Ref.~\onlinecite{santra2008} that the agreement 
with the reference calculations is improved when dispersion interactions are accounted for, shown 
here by results with various dispersion correction schemes.}
\label{fig_vdw_hexam}
\end{figure}

Small water clusters, in particular water hexamers, are an interesting system where dispersion plays a significant role 
(see, e.g. Refs.~\onlinecite{tsai1993hex,xantheas2002,olson2007,santra2008,wang2010}).
The four relevant isomers (known as ``book", ``cage", ``cyclic", ``prism", Fig.~\ref{fig_vdw_hexam}) all have 
total energies that differ by $<$1.3~kJ/mol per molecule according to accurate post-HF methods.\cite{santra2008,klimes2010}
While the ``prism" and ``cage" isomers are preferred by post-HF methods, standard XC functionals find the more open ``cyclic" 
and ``book" isomers to be more stable.
Recent work reveals that dispersion impacts profoundly on the relative energies of the isomers
and improved relative energies are obtained when dispersion is accounted for. 
Indeed the water hexamers have become an important test system with techniques such as DFT-D,\cite{santra2008}  vdW(TS),\cite{santra2008}
vdW-DF,\cite{kelkkanen2009} optB88-vdW,\cite{klimes2010} and the 
modified vdW-DF approach of Silvestrelli (BLYP-vdW(Silv.))\cite{silvestrelli2009} all having been applied.
The performance of these schemes in predicting the relative energies of the water hexamers is shown in Fig.~\ref{fig_vdw_hexam}, where one 
can also see the improvement over a standard functional such as PBE. 
One general point to note, however, is that even without a dispersion correction certain functionals can already give 
quite accurate $absolute$ binding energies for systems held together mainly by 
hydrogen bonding 
and adding dispersion corrections can actually lead to 
too large absolute binding energies. 
This is indeed the case for the water hexamers where, 
for example, the vdW(TS) correction to PBE gives binding energies for the hexamers that are too large by $\approx$4~kJ/mol per molecule.

\subsection{Molecular crystals -- solid benzene}

Molecular crystal polymorph prediction is another important area where dispersion forces can play 
a critical role, and even for molecular crystals comprised of small molecules several polymorphs often 
exist within a small energy window.\cite{neumann2008,price2009}
Identifying the correct energetic ordering of the polymorphs can therefore be a stringent test for any method.
A large number of molecular crystals have been characterised 
experimentally (see, e.g. Ref.~\onlinecite{todorova2010}) and this, therefore, 
could serve as a rich testing ground for DFT-based dispersion schemes. 

Solid benzene is one of the most widely examined test systems, with studies focussing 
on the experimental density and the cohesive energy of the crystal.\cite{meijer1996}
The latter is obtained from the experimental sublimation energy from which the effects of temperature and quantum
nuclear effects must be subtracted giving a value in the 50 to 54~kJ/mol per molecule range.\cite{li2010rpa}
Calculations with PBE give an abysmal cohesive energy of only 10~kJ/mol per molecule and a 
volume $\approx$30\% too large compared to experiment.\cite{bucko2010} 
This huge difference between theory and experiment suggests that dispersion is important to the binding 
of the crystal and indeed the error is greatly reduced when even rather simple schemes are used. 
For example, PBE-D2 and the DCACP potentials give estimates of the lattice energy
(55.7~kJ/mol\cite{bucko2010} and 50.6~kJ/mol,\cite{tapavicza2007} respectively) and densities that are within 10\%
of experiment. 
While the PBE-vdW(TS) scheme overbinds slightly (66.6~kJ/mol), it has been suggested that 
the many-body interactions are quite important for this system and their inclusion reduces 
the PBE-vdW(TS) value by 12~kJ/mol.\cite{tkatchenko2011} 
Non-local vdW-DF overestimates the cell volume by $\approx$10\%\ but gives a rather good cohesive energy of 58.3~kJ/mol.\cite{klimes_unp}
vdW-DF2 reduces both the binding energy (to 55.3~kJ/mol)
and the error in the cell volume, which turns out to be $\approx$3\% larger than the experimental value.\cite{klimes_unp}
Using functionals proposed to improve upon the overly repulsive behavior of vdW-DF, such as optB88-vdW leads to an improvement 
of the reference volume ($\approx$3\%\ smaller than experiment), but the cohesive energy is overestimated, being 69.4~kJ/mol.\cite{klimes_unp}
Again, the overestimated cohesive energy may result from missing many-body interactions. 
The RPA has also been applied to this system and while the density is in very good agreement with experiment,
the cohesive energy is slightly underestimated at 47~kJ/mol.\cite{li2010rpa}
Overall, the improvement over a semi-local functional such as PBE is clear but more applications and tests on
a wider range of systems are needed to help establish the accuracy of the methods.

\begin{figure}[t]
\centerline{
\includegraphics[height=6cm]{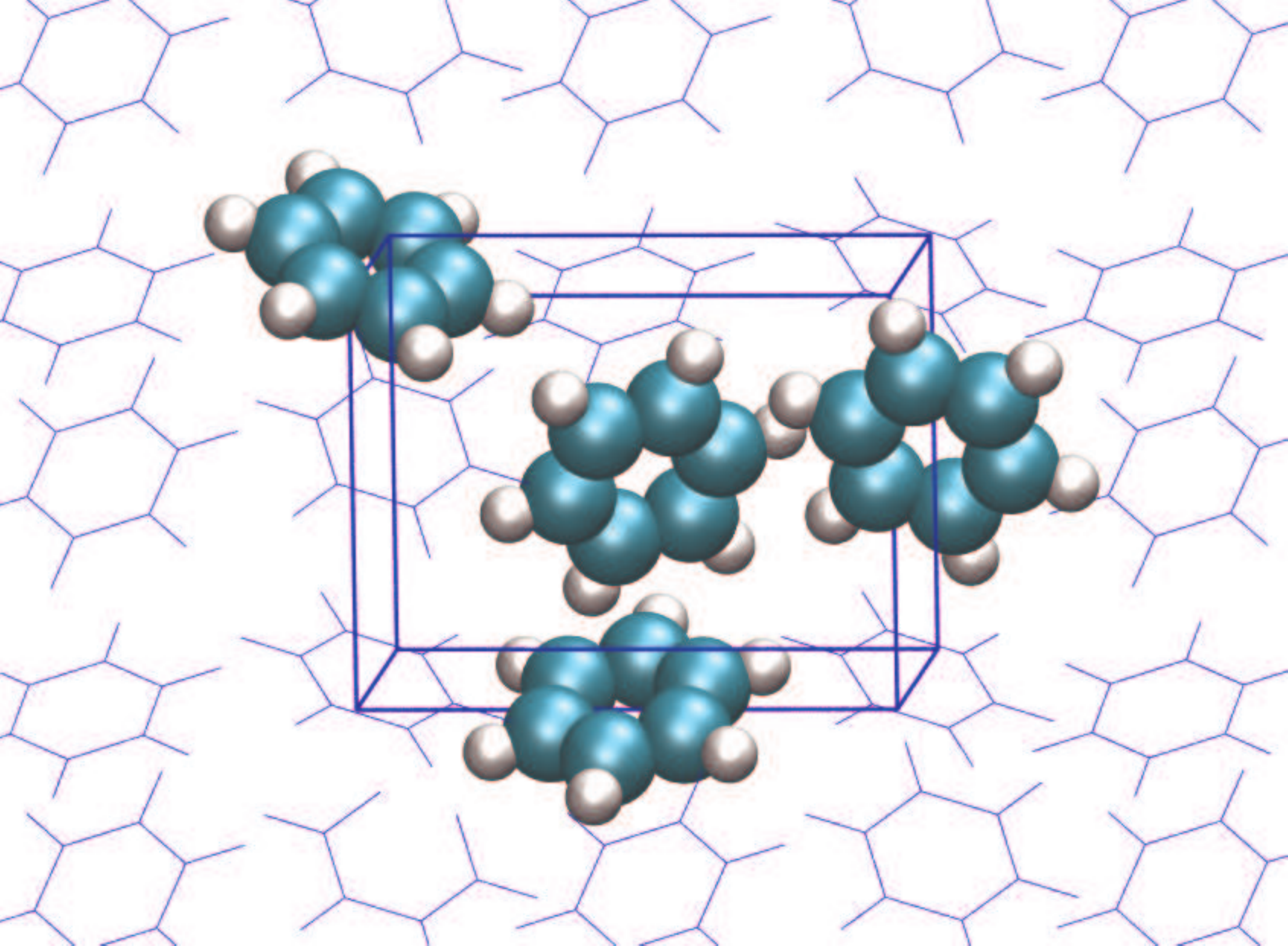}}
\caption[Benzene crystal]
{In molecular crystals the molecules can be packed in a range of orientations and at a range of 
distances from each other. This makes molecular crystals a more challenging class of system than gas phase clusters, 
and, indeed, even for the relatively simple case of benzene shown here, few methods are able to very accurately 
describe the cohesive properties of the crystal.
The benzene molecules in the central unit cell are shown with large balls, all other benzenes in the
neighboring unit cells are shown in wireframe.}
\label{fig_vdw_benzene}
\end{figure}

\subsection{Adsorption -- benzene on Cu(111)}

Adsorption on solid surfaces is another area where great strides forward have recently 
been made with regard to the role of dispersion.
Dispersion is of obvious importance to the binding of noble gases to surfaces but it can also be important
to chemisorption\cite{johnston2008} and, e.g., water adsorption.\cite{carrasco2011}
Indeed for weakly chemisorbed systems dispersion forces attain an increased relative importance
and one can estimate from various 
studies (e.g., Refs.~\onlinecite{thonigold2010} and~\onlinecite{rubes2010}) that dispersion contributes 
about 4 to 7 kJ/mol to the adsorption energy of a carbon-sized atom; not a negligible contribution.

Of the many interesting classes of adsorption system, organic molecules on metals 
have become a hot area of research and for such systems dispersion must be included
if reasonable adsorption energies and structures are to be obtained.
Benzene on Cu(111) is an archetypal widely studied system for organic adsorbates on metals.
It is also an interesting system because it illustrates how difficult it is sometimes to use experimental reference data, 
which for this system has proved to be somewhat of a ``moving target''.
Indirect estimates obtained from temperature programmed
desorption initially placed the adsorption energy at 57~kJ/mol,\cite{lukas2001} however, a recent reinterpretation of the experiment moved
the adsorption energy up to the 66 to 78~kJ/mol range.\cite{ruiz2012}
When looking at this system with DFT, not unexpectedly, PBE gives very little binding (5~kJ/mol).\cite{toyoda2009} 
Most of the step 1 and 2 methods overestimate the binding,\cite{mcnellis2009} for example, PBE-D2 and PBE-vdW(TS)
give 97 and 101~kJ/mol, respectively.
The exception is the work of Tonigold and Gro{\ss}\cite{thonigold2010} where a value of 59~kJ/mol was obtained, 
based on dispersion coefficients obtained by fitting to post-HF data of small clusters.
Recently, Ruiz~{\it et al.}\cite{ruiz2012} approximately included the many-body effects in calculating the $C_6$ 
coefficients in the vdW(TS) scheme, which reduced the predicted adsorption energy to 88~kJ/mol.
vdW-DF gives adsorption energies of about $\sim$53~kJ/mol which underestimates both the old and new reference
data.\cite{toyoda2009,berland2009}
It is clear from this and other systems that adsorption is very challenging for dispersion-based DFT methods at present. 
The ``DFT-D" methods face the problem of obtaining the $C_6$ coefficients for the atoms within the surface of the solid and the 
pairwise methods neglect many-body effects. 
Both of these issues will require much more consideration in the future.

\section{Final remarks}

An enormous amount of progress has been made over the last few years with the treatment 
of dispersion forces within DFT.
It is now one of the most exciting and thriving areas of development in modern computational materials
science and an array of methods has been developed. 
Here we have introduced and classified some of the main, often complementary, approaches. 
We have also discussed how some schemes perform on a selection of systems, including gas phase clusters, 
a molecular solid and an adsorption problem. 
These examples, and the many others in the literature, demonstrate that the range
of systems which can now be treated with confidence with DFT has been greatly extended.
Connected with this, the variety of systems and materials to which dispersion forces are now thought 
to be relevant has grown substantially.
As a result the mantra that ``dispersion forces are not important" is heard less often now and there is 
much less of a tendency to sweep dispersion forces under the rug.

We have seen how many methods have been applied and tested on gas phase molecular clusters, 
for which there are now several approaches that can yield very high accuracy. 
However, a key contemporary challenge is the need to develop methods that will be accurate for both gas phase 
molecular systems and problems involving condensed matter such as adsorption. 
Here, much work remains to be done with regard to the development of DFT-based dispersion techniques for
condensed matter as well as in simply better understanding how current techniques perform in e.g. adsorption
systems where there may be strong polarization effects. 
In this regard, approaches based on the ACFDT such as RPA look promising. 
However, given their high computational cost and complex set-up it is likely that for the foreseeable future such methods will
mainly be useful for \emph{tour de force} reference style calculations rather than for routine studies.\cite{carrasco2012nmat}

Looking to the future, the efficient description of many-body correlation effects in metals and other solids is an important 
unresolved issue. 
Even simply better understanding their importance for different systems would be useful.
For example, although many body correlation effects should be important
in solids, the vdW-DF method which neglects them can perform surprisingly well for solids.\cite{klimes2011}
Another problem closely related to dispersion is the issue of screening which differs substantially in solids and molecules.
For molecules, approaches that neglect screening of exchange, e.g., the so-called long-range corrected (LC) exchange functionals\cite{iikura2001} 
are beneficial. 
Indeed, the LC functionals improve many properties of molecules such as electrostatic moments which in turn decreases an important
source of errors in vdW bonded systems.\cite{sato2007,chai2009,vydrov2009jcp}
In solids, especially metals or semiconductors, the interaction of electrons is significantly modified by the presence of the
other electrons and it is essential to capture this XC effect.
Indeed, no screening of the Fock exchange leads, for example, to overestimated band gaps of semiconductors.\cite{paier2006}
The effect is also significant for correlation, for example, when two homogeneous electron gas spheres are taken from vacuum 
into a homogeneous electron gas background their dispersion interaction is reduced to about a fifth of its original value.\cite{goldstein1989,tao2010}
Since screening is system dependant it is unclear how to treat solids and molecules on the same footing in a 
computationally efficient manner.
A further issue, which is often of minor importance but should be accounted for when 
high accuracy is being sought is anisotropy of the dispersion coefficients.
Isotropic dispersion coefficients seem to be a good first approximation since the anisotropy 
for molecules is on the order of 10\%.\cite{gisbergen1995}
Anisotropy can, however, become an issue for highly anisotropic and polarizable objects\cite{kim2006} and 
to establish the overall importance would be useful.
The anisotropy can be included in some schemes, as has been done, for example, in the BJ model and 
the vdW(TS) approach.\cite{krishtal2009,tkatchenko2011}

Many factors have prompted the recent progress with DFT, with one key factor being the parallel development
of post-HF methods. This has provided the accurate reference data which has served to both shine light
on problems with existing XC functionals and against which new methods can be proved.
The fact that some of this reference data has been easily accessible -- such as the
S22 data set\cite{rezac2008} -- has also helped.
However, as stressed above, an important challenge nowadays is to develop methods that
are accurate for solids and for adsorption.
Unfortunately for these systems accurate reference data are scarce and, indeed, urgently needed either 
from experiment (e.g. microcalorimetry\cite{brown1998} for adsorption)
or higher level electronic structure theories (e.g. post-HF methods,\cite{paulus1999,li2008,nolan2009,marsman2009,tosoni2010,voloshina2011} 
approximations of ACFDT, QMC). 
It is encouraging that progress is being made in both of these areas.
It is also very encouraging that condensed phase reference systems are beginning to emerge, such as ice, LiH, water on LiH and water on 
graphene.\cite{manby2006,nolan2009,marsman2009,rubes2009,jenness2010,binnie2011thesis,ma2011wat,voloshina2011,santra2011,grueneis2011,wellendorff2012} 
By tackling these and other reference systems with the widest possible range of techniques we will
better understand the limitations of existing dispersion-based DFT approaches, which will aid the
development of more efficient
and more accurate methods for the simulation of materials in general.

\begin{acknowledgments}

J. K. was supported by UCL and EPSRC through the PhD+ scheme and A. M. by the European Research Council and the
Royal Society through a Royal Society Wolfson Research Merit Award. 
We'd also like to thank Lo\"ic Roch for help with preparing Figure 1 and D. R. Bowler, A. Gr\"{u}neis, F. Hanke, K. Jordan, G. Kresse, \'E. D. Murray, A. Tkatchenko, J. Wu and 
members of the ICE group for discussions and comments on the manuscript.
\end{acknowledgments}

\end{document}